\def\year{2018}\relax
\documentclass[letterpaper]{article} 
\usepackage{aaai18}  
\usepackage{times}  
\usepackage{helvet}  
\usepackage{courier}  
\usepackage{url}  
\usepackage{graphicx}  
\nocopyright
 
\begin{document}
%
\title{Online Abuse of UK MPs in 2015 and 2017: Perpetrators, Targets, and Topics}

\author{Genevieve Gorrell, Mark Greenwood, Ian Roberts,\\
\Large{\textbf{Diana Maynard and Kalina Bontcheva}}\\
\Large{University of Sheffield, UK}\\
\tt{g.gorrell, m.a.greenwood, i.roberts,}\\
\tt{d.maynard, k.bontcheva@sheffield.ac.uk}
}

\maketitle
\begin{abstract}
  Concerns have reached the mainstream about how social media are
  affecting political outcomes. One trajectory for this is the
  exposure of politicians to online abuse. In this paper we use 1.4
  million tweets from the months before the 2015 and 2017 UK general
  elections to explore the abuse directed at politicians. This
  collection allows us to look at abuse broken down by both party and
  gender and aimed at specific Members of Parliament. It also allows
  us to investigate the characteristics of those who send abuse and
  their topics of interest. Results show that in both absolute and
  proportional terms, abuse increased substantially in 2017 compared
  with 2015. Abusive replies are somewhat less directed at women and
  those not in the currently governing party. Those who send the abuse
  may be issue-focused, or they may repeatedly target an
  individual. In the latter category, accounts are more likely to be
  throwaway. Those sending abuse have a wide range of topical
  triggers, including borders and terrorism.
\end{abstract}


\section{Introduction}
\label{sec:intro}

The recent UK EU referendum and the US presidential election, among
other recent political events, have drawn attention to the increasing
power of social media usage to influence major international
outcomes. Such media profoundly affect our society, in ways which are
yet to be fully understood. One particularly unsavoury way in which
people attempt to influence each other is through verbal abuse and
intimidation. In response to this concern, the UK government has
recently announced a review looking at how abuse and intimidation
affect Parliamentary candidates during
elections.\footnote{https://www.gov.uk/government/news/review-into-abuse-and-intimidation-in-elections}
Shadow Home Secretary Diane Abbott has taken a key role in drawing
attention to the phenomenon and speaking out about how it affects her
and her
colleagues.\footnote{https://www.theguardian.com/politics/2017/jul/12/pm-orders-inquiry-into-intimidation-experienced-by-mps-during-election}\footnote{https://www.theguardian.com/politics/2017/feb/19/diane-abbott-on-abuse-of-mps-staff-try-not-to-let-me-walk-around-alone}

There is a broad perception that intolerance, for example religious or
racial, is on the increase in recent years, some associating this with
the election of Donald
Trump.\footnote{https://www.newyorker.com/news/news-desk/hate-on-the-rise-after-trumps-election}\footnote{https://www.opendemocracy.net/transformation/ae-elliott/assemble-ye-trolls-rise-of-online-hate-speech}
In the UK, the outcome of the EU membership referendum, in which the
British public chose to leave the EU, was also associated with the
legitimisation of racist attitudes and an ensuing increased expression
of those
attitudes.\footnote{https://www.theguardian.com/commentisfree/2016/jun/27/brexit-racism-eu-referendum-racist-incidents-politicians-media}
Twitter is a barometer as well as mediator of these mindsets,
providing a forum where users can communicate their message to public
figures with relatively little personal consequence.

In this work we focus on abusive replies to tweets by UK politicians
in the run-up to the 2015 and 2017 UK general elections. The analysis
focuses on tweets using obscene nouns (``cunt'', ``twat'', etc),
racist or otherwise bigoted language, milder insults (``you idiot'',
``coward'') and words that can be threats (``kill'', ``rape''). In
this way, we define ``abuse'' broadly; ``hate speech'', where
religious, racial or gender groups are denigrated, would be included
in this, but we do not limit our analysis to hate speech. We include
all manner of personal attacks and threats. Obscene language more
generally (e.g. ``fucking'', ``bloody'') was not counted as abusive as
it was less likely to be targeted at the politician personally. These
data allowed us to explore the following questions:

\begin{itemize}

\item{What influences the amount of abuse a politician receives?}

\item{What can we learn about those who send abuse?}

\item{What are the topics of concern to those who send abuse?}

\item{What difference do we see in the above between the two time periods studied?}

\end{itemize}

The work presented is part of a new and under-researched field looking
at how abuse and intimidation are directed at our political
representatives online. As politicians increasingly talk about their
reluctance to expose themselves to this abuse and
intimidation\footnote{https://www.theguardian.com/technology/2016/jun/18/vile-online-abuse-against-women-mps-needs-to-be-challenged-now},
we see that there is a very real danger that they may no longer choose
to do this work, and the part they play in creating a fair
representation of the electorate will be lost. For this reason, it is
important to engage with this aspect of the way the web is affecting
our society. Previous work has examined abusive behaviour online
towards different groups, but the reasons why a politician might
inspire an uncivil response are very different to an ordinary member of the public, with resulting different implications. Additionally, to
the best of our knowledge, this is the first study to contrast
quantitative changes across two comparable but temporally distinct
samples (the two general election periods).


\section{Related Work}
\label{sec:related}

Whilst online fora have attracted much attention as a way of exploring
political
dynamics~\cite{nulty2016social,kaczmirek2013social,colleoni2014echo,weber2012mining,conover2011political,gonzalez2010emotional},
and the effect of abuse and incivility in these contexts has been
explored~\cite{vargo2017socioeconomic,rusel2017bringing,gervais2015incivility,hwang2008does},
little work exists regarding the abusive and intimidating ways people
address politicians online; a trend that has worrying implications for
democracy. Theocharis et al~\shortcite{theocharis2016bad} collected
tweets centred around candidates for the European Parliament election
in 2014 from Spain, Germany, the United Kingdom and France posted in
the month surrounding the election. They find that the extent of the
abuse and harrassment a politician is subject to correlates with their
engagement with the medium. Their analysis focuses on the way in which
uncivil behaviour negatively impacts on the potential of the medium to
increase interactivity and positively stimulate
democracy. Stambolieva~\shortcite{stambolieva2017methodology} studies
online abuse against female Members of Parliament (MPs) only; in
studying male MPs as well, we are able to contrast the level of abuse
they each receive. Furthermore, we contrast proportional with absolute
figures, creating quite a different impression from the one she gives.

A larger body of work has looked at hatred on social media more
generally~\cite{bartlett2014antisocial,perry2009cyberhate,coe2014online,cheng2015antisocial}. Williams
and Burnap present work demonstrating the potential of Twitter for
evidencing social models of online hate crime that could support
prediction, as well as exploring how attitudes co-evolve with events
to determine their
impact~\cite{williams2015cyberhate,burnap2015cyber}. Silva et
al~\shortcite{silva2016analyzing} use natural language processing (NLP) to
identify the groups targeted for hatred on Twitter and Whisper.

Work exists regarding accurately identifying abusive messages
automatically~\cite{burnap2015cyber,nobata2016abusive,chen2012detecting,dinakar2012common,wulczyn2017ex}. The
work of Wulczyn et al has been described as the state of the art, with
precision/recall of 0.63 being reported as equivalent to human
performance. Bartlett et al~\shortcite{bartlett2014antisocial} report
a human interannotator agreement of only 0.69 in annotating racial
slurs, and Theocharis et al~\shortcite{theocharis2016bad} report a
human agreement of 0.8 with a Krippendorf's alpha of 0.25 on UK data,
demonstrating that the limiting factor is the complexity of the task
definition. Burnap and Williams~\shortcite{burnap2016us} particularly
focus on hate speech with regards to protected characteristics such as
race, disability and sexual orientation. Waseem and
Hovy~\shortcite{waseem2016hateful} also focus on hate speech, and
share a gold standard UK Tweet corpus. Dadvar et
al~\shortcite{dadvar2013expert} seek to identify the problem accounts
rather than the problem material. Schmidt and
Wiegand~\shortcite{schmidt2017survey} provide a review of prior work
and methods.

In the next section, we describe our data collection methodology. We
then present our results, beginning with an analysis of who receives
the abuse, before moving on to who sends it and the topics that are 
most likely to trigger abusive replies.


\section{Data Collection}
\label{sec:data}

The 2015 corpus was created by downloading tweets in real-time using
Twitter's streaming API. Tweets posted from the end of May 6th
to the end of June 6th (the day before the election) were collected.
The data collection focused on Twitter accounts of MPs, candidates,
and official party accounts. We obtained a list of all current
MPs\footnote{From a list made publicly available by BBC News Labs,
which we cleaned and verified} and all currently known election
candidates\footnote{List of candidates obtained from
https://yournextmp.com} (at that time) who had Twitter accounts (506
currently elected MPs and 1811 candidates, of whom 444 MPs were also
standing for re-election). We collected every tweet by each of these
users, and every retweet and reply (by anyone) for the month leading
up to the 2015 general election. The methodology was the same for the
2017 collection, this time collecting from the end of April 7th to the
end of May 7th, for 1952 candidates and 480 sitting MPs, of whom 417
were also candidates. Data were of a low enough volume not to be
constrained by Twitter rate limits. Corpus statistics are given in
table~\ref{tab:corpus}, and are separated out into all politicians
studied and just those who were then elected as MPs.

\begin{table}
\begin{center}
  \begin{tabular}{|l|l|l|l|}
   \hline
    & \#collected & \#hadabuse & \%abusive\\
   \hline
   2015 MPs+cands & 597 411 & 16 628 & 2.8\%\\
   2015 MPs & 277 000 & 10 091 & 3.6\%\\
   2017 MPs+cands & 821 662 & 32 791 & 4\%\\
   2017 MPs & 613 952 & 24 659 & 4\%\\
 \hline
 \end{tabular}
\caption{Corpus statistics}
\label{tab:corpus}
\end{center}
\end{table}

In order to identify abusive language, the politicians it is targeted
at, and the topics in the politician's original tweet that tend to
trigger abusive replies, we use a set of NLP
tools, combined into a semantic analysis pipeline. It includes, among
other things, a component for MP and candidate recognition, which
detects mentions of MPs and election candidates in the tweet and pulls
in information about them from DBpedia. Topic detection finds mentions
in the text of political topics (e.g. environment, immigration) and
subtopics (e.g. fossil fuels). The list of topics was derived from the
set of topics used to categorise documents on the gov.uk
website\footnote{e.g. https://www.gov.uk/government/policies}, first
seeded manually and then extended semi-automatically to include
related terms and morphological variants using
TermRaider\footnote{https://gate.ac.uk/projects/arcomem/TermRaider.html},
resulting in a total of 940 terms across 51 topics.
We also perform hashtag tokenization, in order to find
abuse and threat terms that otherwise would be missed.
In this way, for example, abusive language is found in the hashtag 
``\#killthewitch''.

A dictionary-based approach was used to detect abusive language in
tweets. An abusive tweet is considered to be one containing one or
more abusive terms from the vocabulary list.\footnote{see
supplementary materials for anonymous review} This contained 404
abuse terms in British and American English, comprising mostly an
extensive collection of insults, with a few threat terms such as
``kill'' and ``die'' also included. Racist and homophobic terms are
included as well as various terms that denigrate a person's appearance
or intelligence.

Data from Kaggle's 2012 challenge, ``Detecting Insults in Social
Commentary''\footnote{https://www.kaggle.com/c/detecting-insults-in-social-commentary/data},
was used to evaluate the success of the approach, comprising two
similar corpora of short online comments marked as either abusive or
not abusive, amounting to around 6000 items. Our approach was shown to
have an accuracy of 0.78 (Cohen's Kappa: 0.39) on the first set, with
a precision of 0.62, a recall of 0.45 and an F1 0.53; and 0.78 on the
second (Cohen's Kappa: 0.36), a precision of 0.60, recall of 0.43 nad
F1 of 0.50. In practice, performance is likely to be slightly better
than this, since the Kaggle corpus is American English. This
performance is comparable to that obtained by
Stambolieva~\shortcite{stambolieva2017methodology}, though somewhat
lower than the state of the art, where F1s in excess of 0.6 have been
reported on longer texts (see the previous section). Given the noisy
and brief nature of tweets, however, it is often the case that deeper
semantic analysis methods, e.g. dependency parsing, tend to perform
poorly, which is why we opted for a robust dictionary-based
approach. Manual review of the errors shows that some false positives
arise in the area of threat terms, such as ``rape'' and ``murder'',
which are often discussed as being topics of concern, and there are
also a small number of cases where strong language is used simply for
emphasis or even meant positively, as in one case where a politician
was praised for ``working his balls off'' and encouraged to ``get his
arse into Downing Street''. (Downing Street is the location of the UK
government headquarters.) The errors do not seem to show any
particular bias that might affect the results reported here.


\section{Who is receiving the abuse?}

\begin{figure*}[t]
\centering
\includegraphics[width=0.8\textwidth]{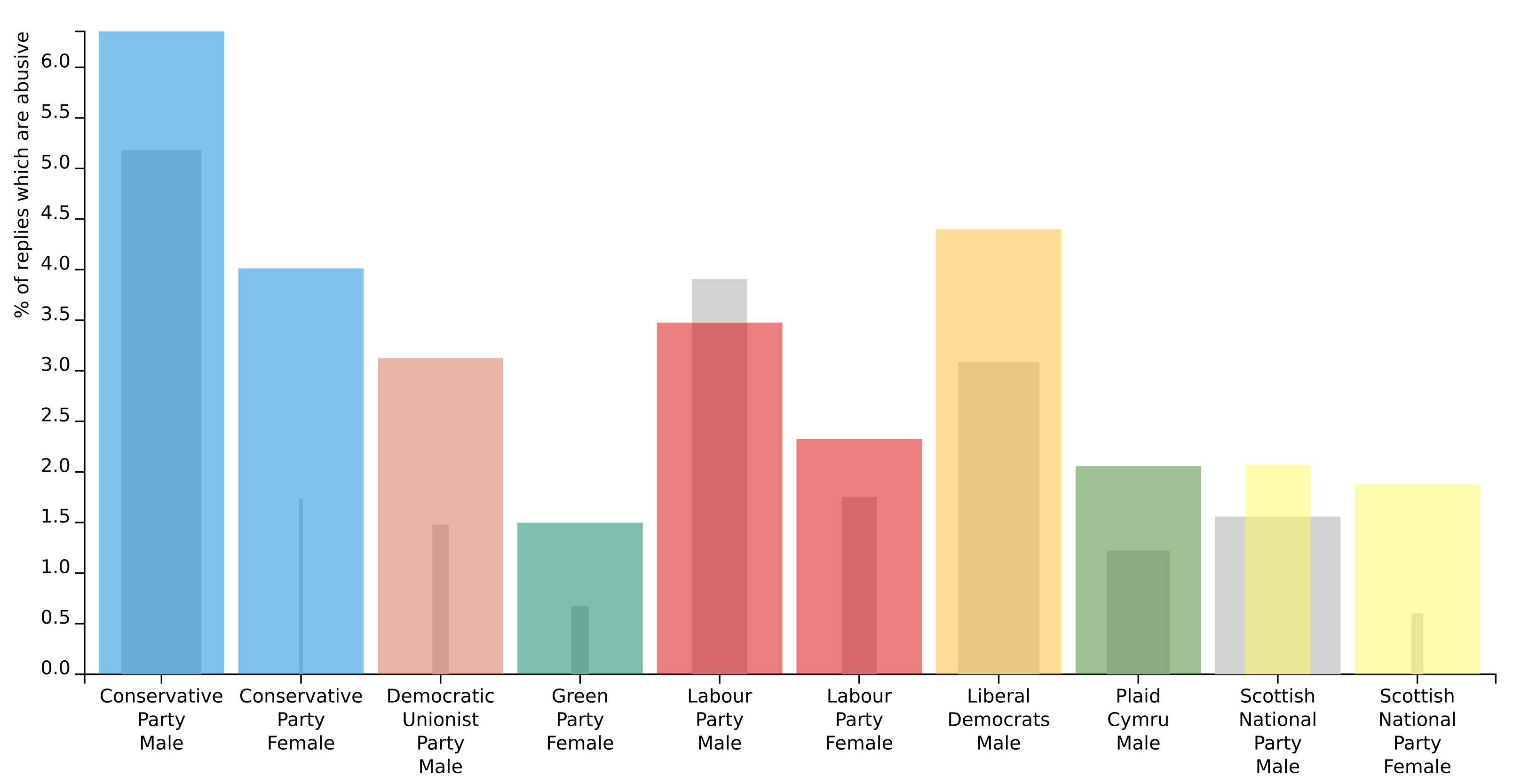}
\caption{Rise in abuse from 2015 (grey bars) to 2017}
\label{fig:riseInAbuse}
\end{figure*}


One claim made by politicians from all parties is that the amount of
abuse, in terms of both volume and proportion, has increased in recent
years. With our dataset spanning both the 2015 and 2017 UK general
elections, we have the opportunity to contribute some evidence on this
point. Figure \ref{fig:riseInAbuse} shows both the proportion of
replies which are abusive (the heights of the bars) for 2015 (grey
bars) and 2017 (bars coloured by party), and the change in volume of
abusive replies (the width of the bars). The graph shows that in most
cases, regardless of party or gender (note that only party/gender
combinations which are relevant to both 2015 and 2017 are shown), both
the proportion and volume have increased between 2015 and 2017. The
two exceptions are where the proportion of abusive replies has fallen
for male Labour Party MPs, and where the volume of abusive replies has
fallen for male Scottish National Party (SNP) MPs. The former is difficult to explain, whereas the latter can be easily attributed to the loss of seats (the party as a whole went from winning 56 seats in 2015 to winning just 35 in 2017).
In most cases,
however, it is clear that both the proportion and volume of abusive
replies have risen in the two years between elections. In some cases
this rise has been dramatic, noticeably for female Conservative MPs,
although there are clear reasons (Theresa May becoming Prime Minister
in this case) behind most of the dramatic changes. However, these are
better investigated through other analyses and visualizations.

\begin{figure}[t]
\includegraphics[width=\columnwidth]{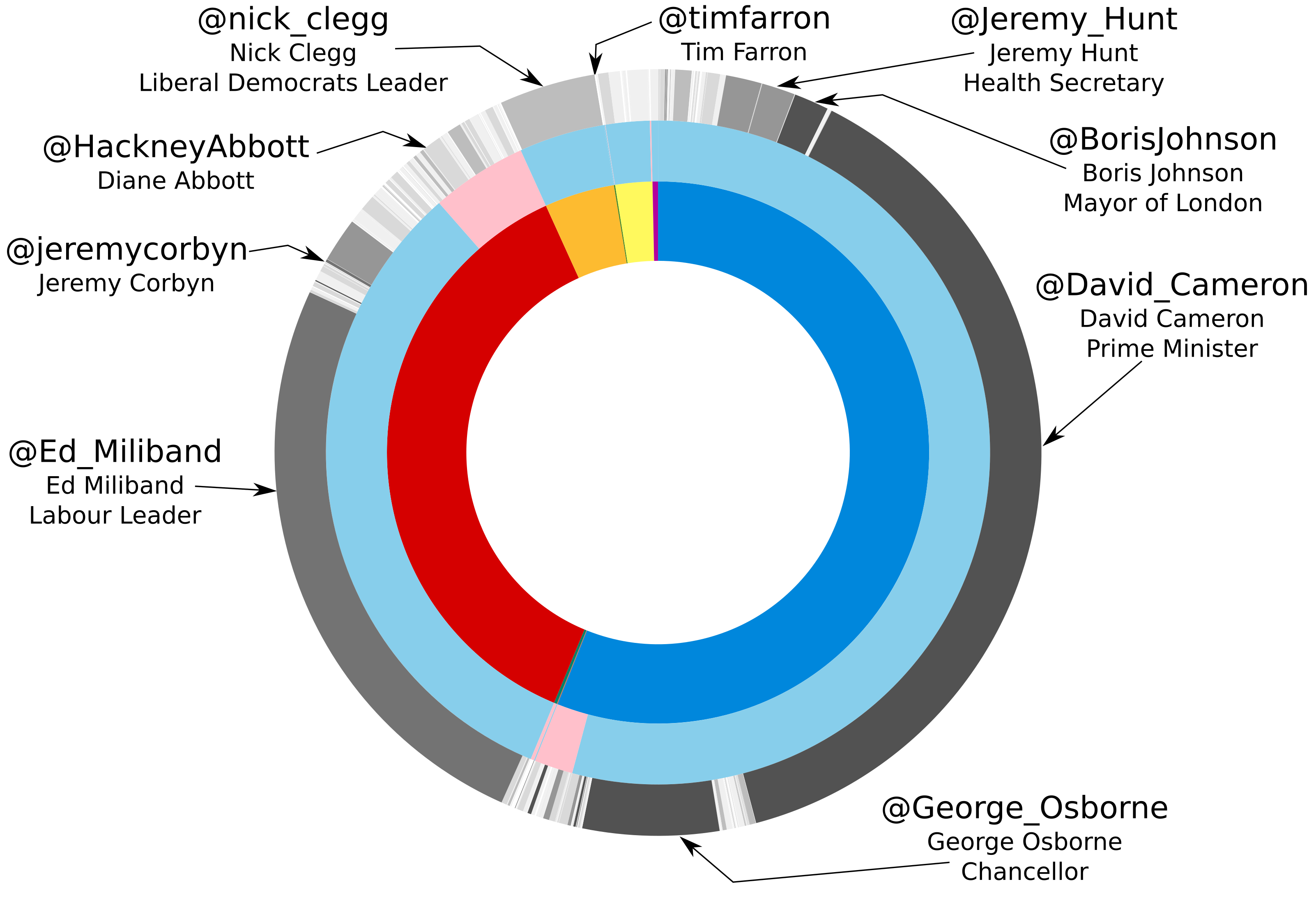}
\caption{Abuse per MP in 2015}
\label{fig:abuse2015Proportion}
\end{figure}

\begin{figure}[t]
\includegraphics[width=\columnwidth]{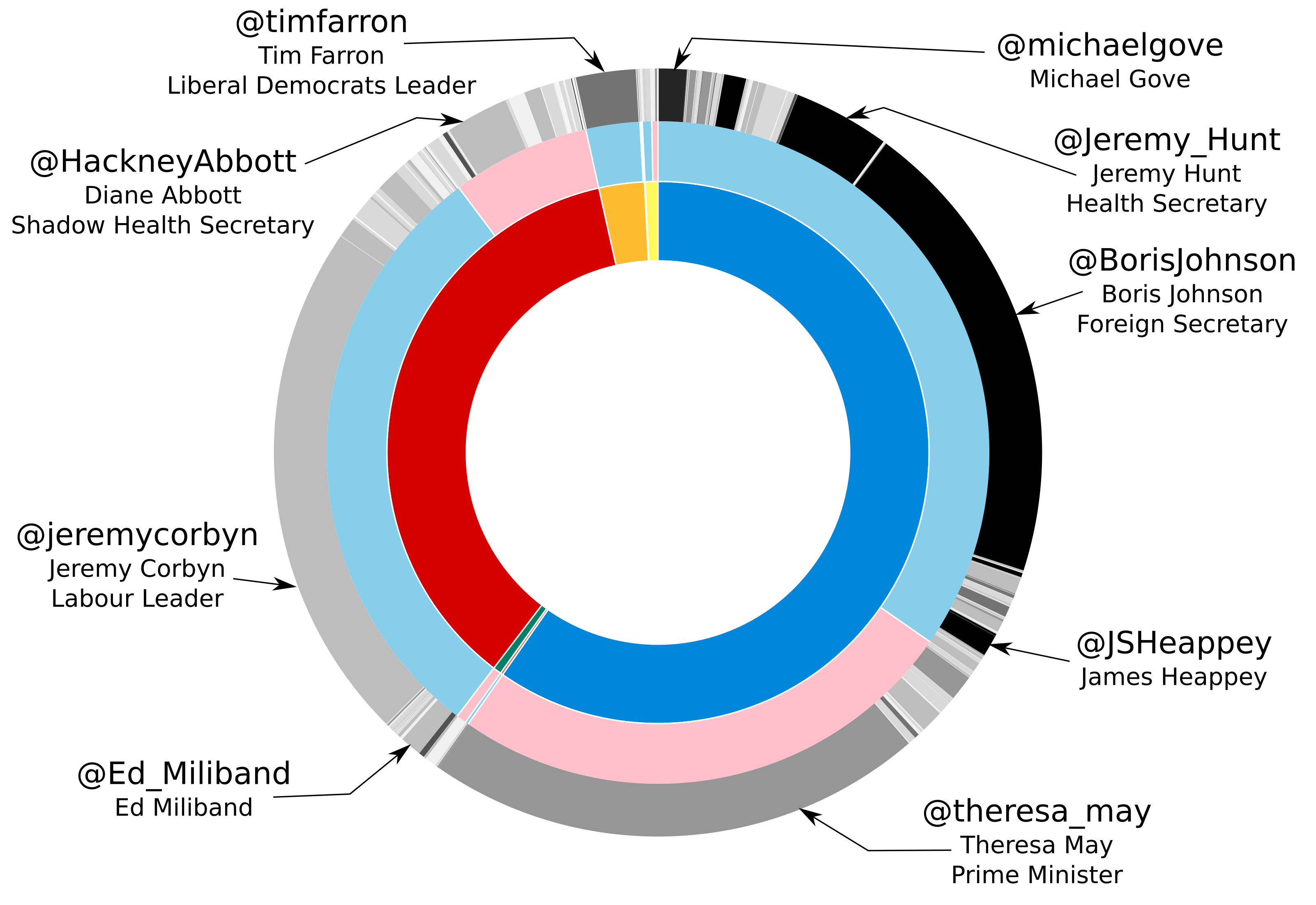}
\caption{Abuse per MP in 2017}
\label{fig:abuse2017Proportion}
\end{figure}

In figures \ref{fig:abuse2015Proportion} and
\ref{fig:abuse2017Proportion}, also available online in interactive
form\footnote{http://greenwoodma.servehttp.com/data/buzzfeed/sunburst.html},
the outer ring represents MPs, and the size of each outer segment is
determined by the number of abusive replies each receives. Working
inwards you can then see the MP's gender (pink for female and blue for
male) and party affiliation (blue for Conservative, red for Labour,
yellow for the Scottish National Party -- see the interactive graph to
explore the smaller parties). The darkness of each MP segment denotes
the percentage of the replies they receive which are abusive. This
ranges from white, which represents up to 1\%, through to black, which
represents 8\% and above. It is clear that some MPs receive
substantially more abusive replies than others. These visualizations
highlight that prominence seems to be a bigger indicator of the
quantity of abusive replies an MP receives than party or gender. In
2015 the majority of the abusive replies were sent to the leaders of
the two main political parties (David Cameron and Ed Miliband). In
2017 the people receiving the most abuse has changed, but again the
leaders of the two main parties (Jeremy Corbyn and Theresa May)
receive a large proportion of the abuse, followed by other prominent
members of the two parties. Unlike in 2015 though, in 2017 there are
other MPs who receive a large proportion of the abuse; most notably
Boris Johnson, who rose to national prominence as one of the main
proponents of the Leave campaign during the UK European Union
membership referendum and with his subsequent promotion to Foreign
Secretary. Conversely, Ed Miliband has seen the volume of abuse he
receives reduce dramatically with his return to the back benches after
quitting as leader of the Labour Party after the 2015 election.

Whilst this view is useful in understanding the amount of abuse
politicians receive in absolute terms, it is less valuable in
illustrating the different responses the parties and genders receive,
because individual effects dominate the picture. Therefore it is
important to also see the results per-MP. The online version of these
graphs includes a ``count'' view in which each segment of the outer
ring represents a single MP. It is evident at a glance that replies to
male Conservative MPs are proportionally more abusive, with female
Conservative MPs not far behind, an impression that will be
further explored statistically below.

\begin{figure}[t]
\includegraphics[width=\columnwidth]{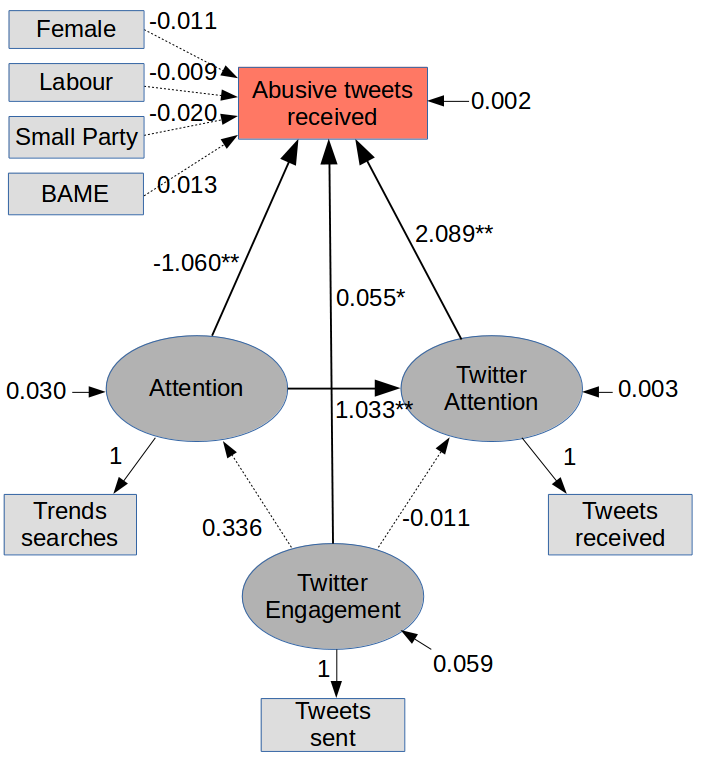}
\caption{Abuse per MP in 2017}
\label{fig:sem}
\end{figure}

Structural equation modeling (see Hox and
Bechger~\shortcite{hox2007introduction} for an introduction) was used
to broadly relate three main factors with the amount of abuse
received: prominence, Twitter prominence (which we hypothesise differs
from prominence generally) and Twitter engagement. We obtained Google
Trends data for the 50 most abused politicians in each of the time
periods, and used this variable as a measure of how high-profile that
individual is in the minds of the public at the time in
question. Search counts for the month running up to each election were
totalled to provide a figure. We used number of tweets sent by that
politician as a measure of their Twitter engagement, and tweets
received as a measure of how high-profile that person is on
Twitter. The model in figure~\ref{fig:sem}, in addition to proposing
that the amount of abuse received follows from these three main
factors, also hypothesises that the amount of attention a person
receives on Twitter is related to their prominence more generally, and
that their engagement with Twitter might get them more attention, both
on Twitter and beyond it. It is unavoidably only a partial attempt to
describe why a person receives the abuse they do, since it is hard to
capture factors specific to that person, such as any recent
allegations concerning them, in a measure. The model was fitted using
Lavaan,\footnote{http://lavaan.ugent.be/} resulting in a chi-square
with a p-value of 0.403 (considered satisfactory, see Hox and
Bechger~\shortcite{hox2007introduction}), and shows a number of
significant findings (indicated with a bold line and asterisks against
the regression figure). A strong pathway to receiving more abuse on
Twitter is simply that if a person is well-known, they receive a lot
of tweets, and if they receive a lot of tweets, they receive a lot of
abusive tweets, in absolute terms. However, an additional pathway
shows that having removed this numbers effect from consideration,
being very well known (if Google searches can be taken as a measure of
that) leads to a person being \textit{less} likely to receive abuse on
Twitter. Perhaps to certain senders of abuse, a large target is a less
attractive one. A further pathway positively relates Twitter
engagement to abuse received, supporting Theocharis et
al's~\shortcite{theocharis2016bad} findings. In this case, perhaps it
is what the person said that provides an attractive target for
abuse. The suggestion is that there may be different types of abuse
going on.

The effects that aren't significant are also interesting. For example,
engaging more with Twitter doesn't correlate with getting more
attention on Twitter. The impact of gender, party and ethnicity is,
though somewhat telling, uncompelling in this model, so these are
explored separately. In the model, being female and being a Labour
politician emerge as factors tending to reduce abuse received,
relative to being a male or a Conservative; being a member of any
other party even more so. Being an ethnic minority may tend to
increase it. Note however that the ethnicity data is sparse, so
unlikely to reveal a significant result.

In the full set of 2017 politicians, male MPs received around 2
abusive tweets in every 100 responses ($\sigma=.0261$), compared with
women MPs' 1.3 ($\sigma=.0139$). (Note that these numbers appear low
compared with the overall corpus statistics because of the long tail
of MPs with little abuse.) An independent samples t-test finds the
difference significant ($p\textless.001$).  In order to test the
possibility that the effect arises from a small number of dominant
individuals, we removed those politicians receiving in excess of
10,000 tweets in the sample period, namely Angela Rayner, Boris
Johnson, Diane Abbott, Jeremy Hunt, Jeremy Corbyn, Tim Farron and
Theresa May.  The difference remains significant ($p\textless.01$)
even after the removal of these outliers. This result is also found in
the 2015 data ($p\textless.05$), remaining significant though weaker
even after the removal of, in the case of 2015, three male outliers;
David Cameron, Ed Milliband and Nick Clegg. In 2015, as discussed
above, the level of abuse was lower, with men receiving around 1.3
abusive tweets in every 100 ($\sigma=.0131$) to women's 1
($\sigma=.0105$). Black, Asian and ethnic minority (BAME) MPs received
marginally less abuse than non-BAME MPs in the 2017 sample (1.4\% as
opposed to 1.7\%; data not available for 2015), but the result was not
significant, and contradicts the finding above, suggesting that other
factors may account for it. Recall also that the BAME sample size is
small.

Differences in the proportion of abusive tweets received across
parties are also apparent. In an independent samples t-test,
Conservative MPs received significantly more abusive tweets than
Labour ($p\textless.001$) in the time frame studied, with an average of
2.3 percent ($\sigma=.0286$) as opposed to 1.3 percent
($\sigma=.0135$). This result is unaltered by the removal of
the above outliers; Conservatives received 2.2 abusive tweets per
hundred as opposed to 1.2 by Labour MPs, and the significance of the
result remains the same. Conservatives also receive more abuse in
2015; however, in 2015 the result is not significant, perhaps
suggesting a greater dependence on the prevailing political
circumstances. It would certainly be interesting to review whether
this result changes should the Conservatives stop being the governing
party.




\section{Who is sending the abuse?}

Having established which politicians tend to be targeted by abusive
messages, now let us examine the accounts behind these posts. For this
analysis we focused on the 2506 Twitter accounts in our 2017 dataset
who have sent at least three abuse-containing tweets. A random sample
of 2500 tweeters for whom we found no abusive tweets were then
selected to form a contrast group.

Independent samples t-tests revealed that those who tweeted abusively
have more recent Twitter accounts by a few months (1533 days on
average vs 1608, $p\textless.001$), smaller numbers of favourited tweets
(7379 vs 14596, $p\textless.001$), fewer followers (1085 vs 3260,
$p\textless.05$), follow fewer accounts (923 vs 1472, $p\textless.05$),
are featured in fewer lists (23 vs 67, $p\textless.001$) and have fewer
posts (16445 vs 25258, $p\textless.001$). After partialing out account
age, number of abusive tweets still correlated significantly with
number of favourites ($p\textless.001$), number of followed accounts
($p\textless.001$), number of times listed ($p\textless.01$) and number of
posts ($p\textless.001$), demonstrating that with the exception of
follower number, these relationships cannot be explained by account
age. One explanation for these findings would be that a certain number
of accounts are being created for the purpose of sending anonymous
abuse.

To investigate in more detail the abuse-sending behaviour of these
2,506 accounts, we define a new metric, ``targetedness'', in order to
differentiate between those users who send strongly worded tweets to a
wide number of politicians (perhaps because their strength of feeling
pertains to an issue rather than a particular person to whom they
tweet) and those users that target a particular individual or a small
number of politicians.  The metric $t$ is given in
equation~\ref{eqn:targetedness}. It divides the total number of
abuse-containing tweets found in our sample authored by that Twitter
user, $a$, by the number of separate recipients to whom they were
directed, $r$. Multiplying the divisor by two and subtracting one (a
conservative choice that yet achieves the objective) provides a simple
way to deflate the score for those with more recipients, compared with
those achieving the same ratio with a smaller number of tweets to a
smaller number of people. In this way, for example, sending five
abuse-containing tweets to five different politicians results in a
lower targetedness score than sending one abuse-containing tweet to
one politician. The metric has the advantage (compared with for
example Gini coefficient or entropy) of appropriately positioning the
``long tail'' of accounts with little abuse in the middle ground.

\begin{eqnarray} \label{eqn:targetedness}
  t = \frac{a}{(2*r)-1)}
\end{eqnarray}

This metric was used to split the group into three parts: those with a
score higher than one ($n=708$), whose abuse pattern we refer to as
``targeted''; those with a score lower than one ($n=646$), described
below as ``broad''; and those scoring one ($n=1021$), which we dub
``responsive''. These three groups show distinct behavioural
patterns. For example, those that target one or two individuals for
repeated abusive tweets (``targeted'') include one Twitter user who
addressed Jeremy Hunt with a single word epithet 28 times in our
sample, and one account, now deleted, who made MP James Heappey the
focus of their exclusive uncivil attention to the tune of 34
tweets. James Heappey made an unpopular comment to a schoolchild
shortly before the 2017 general election, which might have been a
contributing factor in the attention he received. Those that send a
large number of abuse-containing tweets thinly spread among a large
number of politicians (``broad'') tend to be ideologically driven; for
example, one user addresses ten different individuals and focuses
consistently on class war.

\begin{figure}
  \caption{Av. Account Ages (Days)}
  \centering
    \includegraphics[width=0.8\columnwidth]{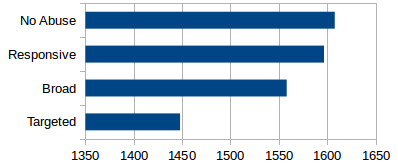}
\label{fig:ab-acc-age}
\end{figure}

\begin{figure}
  \caption{Av. Following/Followers}
  \centering
    \includegraphics[width=0.8\columnwidth]{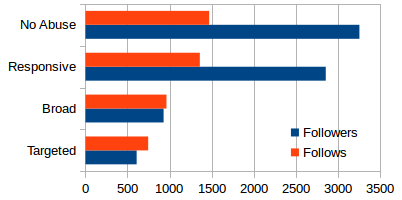}
\label{fig:ab-fol}
\end{figure}

\begin{figure}
  \caption{Average Number of Posts}
  \centering
    \includegraphics[width=0.8\columnwidth]{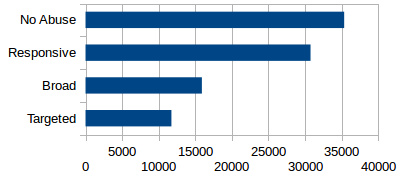}
\label{fig:ab-posts}
\end{figure}

Both targeted and broad users show indicators that differentiate them
to a greater extent from tweeters who were not found to have sent any
abusive tweets; namely, the former have fewer posts, fewer favourites,
newer accounts, fewer followers and fewer followees, as well as
appearing on fewer lists, with the ``targeted`` group more pronounced
in these effects than the ``broad'' group. There are also
abuse-sending tweeters in the middle ground -- the
``responsive''. These address a medium number of people with a medium
amount of abuse and have account statistics that are closer to normal
(though note that the inclusion of the long tail of accounts with
little abuse in the ``responsive'' category may contribute to
this). Figures~\ref{fig:ab-acc-age}, \ref{fig:ab-fol} and
\ref{fig:ab-posts} illustrate these findings, showing average account
age, average number of followed accounts, average number of followers
and average number of posts respectively.

Only in the group showing targeted abuse behaviour do we see a
significant difference in account age compared with the control group
(1448 days on av. vs 1606 with a higher standard deviation; 1208 days
vs 1023, $p\textless.001$). The other two types of abuse profile do
not show a significant effect with account age, perhaps indicating
that throwaway accounts are more likely to be occurring in the former
group. All other indicators for all abuse types remain significantly
different from the control group for whom no abuse was found, other
than follower number.

In order to establish that number of abusive tweets alone does not
explain the effect, a sample was selected of tweeters with 3 to 9
abusive tweets inclusive. This resulted in a lower average number of
abusive tweets from the ``targeted'' group ($n=519$) compared with the
``broad'' group ($n=674$). Even within this sample, the ``targeted''
group had substantially younger Twitter accounts, fewer favourites,
fewer followers, followed fewer accounts, were listed less often and
had posted less.

\begin{figure*}
  \caption{Account Ages}
  \centering
  \includegraphics[width=0.95\textwidth]{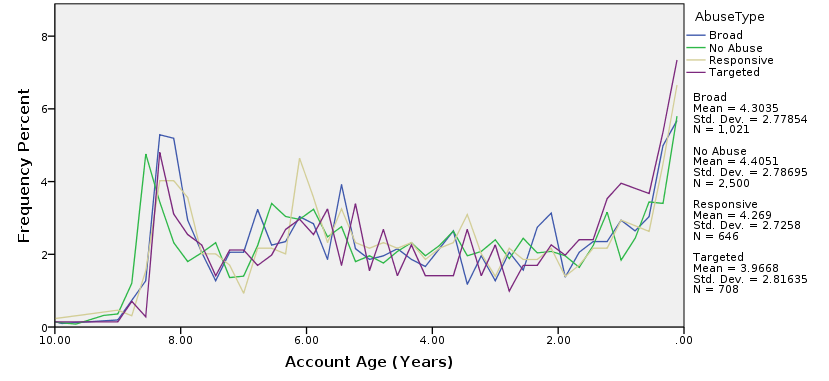}
\label{fig:timeline}
\end{figure*}

In Figure~\ref{fig:timeline} we plot the account age in years against
the percentage of accounts from each category with that age, allowing
us to explore the differences in account age between the different
abuse types in more detail. Amongst all 8 to 10 year old accounts,
those inclined to send abuse were created later compared with those
that did not. In other words, the oldest of all non-abuse accounts are
a little older than the oldest of the accounts that send
abuse. Secondly, in the most recent 1.5 years, many Twitter accounts
have been created, including a prominent spike even for the control
group. This indicates a fairly high rate of account creation, but this
effect is more pronounced for those that send abuse, particularly
targeted abuse, perhaps suggesting that the intention to abuse is a
prominent reason, though not the only one, for creating a throwaway
account.

A similar analysis of accounts that sent abusive tweets in the leadup
to the 2015 general election revealed some differences compared with
2017. Firstly, whilst abuse-posting accounts were again younger, in
2015 they posted more (12732 statuses on average vs 8752,
$p\textless.001$) and favourited more (2499 vs 1517,
$p\textless.001$). Reviewing the data reveals that in 2015 more abuse
was sent by a smaller number of individuals, including a substantial
number of what might be termed serial offenders, who perhaps
explicitly seek attention by posting and favouriting. The greater
quantity of abuse found in the 2017 data is more thinly spread across
a larger number of lesser offenders. Twitter's commitment to reviewing
and potentially blocking abusive users might account for the
difference in the two years. It is striking that despite evident
progress in that regard, the amount of abuse has still
increased. Manual review of the data shows no evidence of ``bots''
(automated accounts) in the sample. Though bot activity is common in
Twitter political contexts~\cite{kollanyi2016bots}, we suggest that
bots are unlikely to use abusive language.

In both the 2015 dataset and the 2017 one we found that
significantly more accounts had been closed from the group that sent
abusive tweets; 16\% rather than 6\% ($p\textless.01$) in 2015 and 8\%
rather than 2\% ($p\textless.001$) in 2017 (Fisher's exact test).

\section{Topics Triggering Abusive Replies}

To explore what motivates people to send abusive tweets, we begin by
analysing what topics are mentioned by them in their abusive responses
to candidates. Abusive tweets were compared against the set of
predetermined topics described earlier. Words
relating to these topics were then used to determine the level of
interest in these topics among the abusive tweets, in order to gain an
idea of what the abusive tweeters were concerned about. So for
example, the following 2017 tweet is counted as one for ``borders and
immigration'' and one for ``schools'': \textit{``Mass immigration is
  ruining schools, you dick. We can't afford the interpretation
  bill.''}

\begin{figure}
  \caption{Topics in Abusive Responses 2017}
  \centering
    \includegraphics[width=\columnwidth]{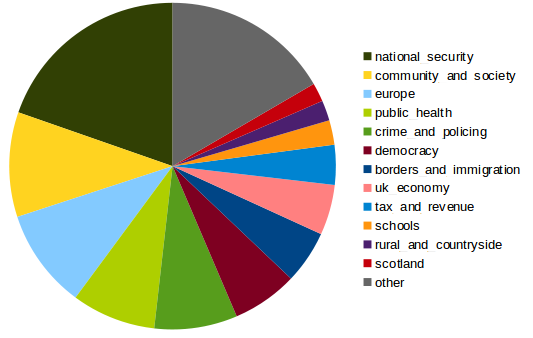}
\label{fig:topics-in-abuse-2017}
\end{figure}

\begin{figure}
  \caption{Topics in All Tweets 2017}
  \centering
    \includegraphics[width=\columnwidth]{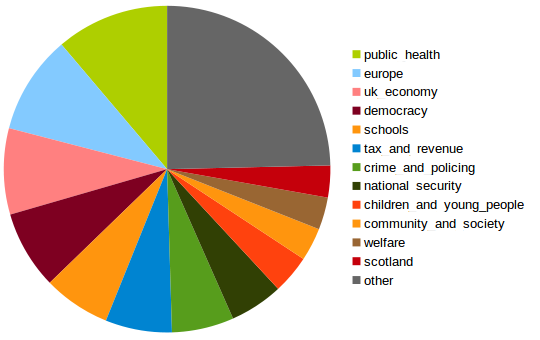}
\label{fig:topics-in-background-2017}
\end{figure}

\begin{figure}
  \caption{Topics in Abusive Responses 2015}
  \centering
    \includegraphics[width=\columnwidth]{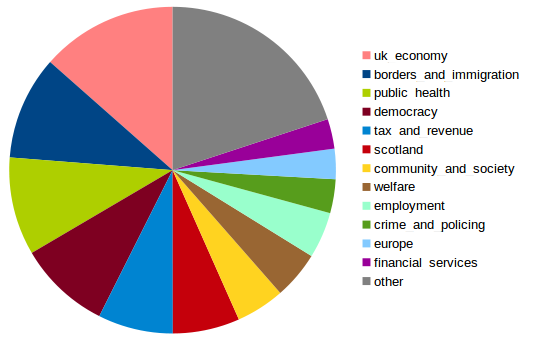}
\label{fig:topics-in-abuse-2015}
\end{figure}

\begin{figure}
  \caption{Topics in All Tweets 2015}
  \centering
    \includegraphics[width=\columnwidth]{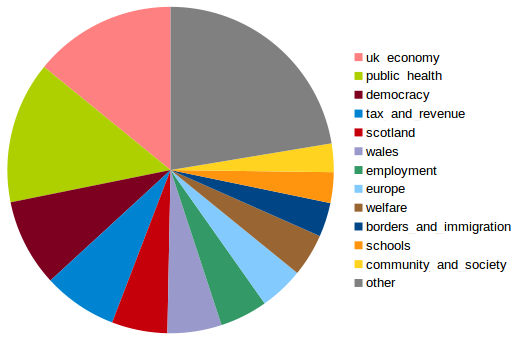}
\label{fig:topics-in-background-2015}
\end{figure}

The topic titles shown in 
figures~\ref{fig:topics-in-abuse-2017},
\ref{fig:topics-in-background-2017},~\ref{fig:topics-in-abuse-2015},
~\ref{fig:topics-in-background-2015} 
are generally self-explanatory, but a few require
clarification. ``Community and society'' refers to issues pertaining
to minorities and inclusion, and includes religious groups and
different sexual identities. ``Democracy'' includes references to the
workings of political power, such as ``eurocrats''. ``National
security'' mainly refers to terrorism, where ``crime and policing''
does not include terrorism. The topic of ``public health'' in the UK
is dominated by the National Health Service (NHS). ``Welfare'' is
about entitlement to financial relief such as disability allowance and
unemployment cover.

In particular, figure~\ref{fig:topics-in-abuse-2017} gives mention
counts for these topics in abuse-containing tweets posted in response
to politicans' tweets in the month leading up to the 2017 general
election, and respectively, figure~\ref{fig:topics-in-abuse-2015} does
so for 2015. As a comparison,
figure~\ref{fig:topics-in-background-2017} shows the topics mentioned
in all tweets in the same month for 2017, and
figure~\ref{fig:topics-in-background-2015} -- those for 2015.

It is notable that in 2017 national security comes to the fore in
abuse-containing tweets, whilst only being the eighth most prominent
topic in the whole set of tweets. Similarly, community and society is
more frequent in abuse-containing tweets than in tweets in general. In
the month preceding the 2017 election, the UK witnessed its two
deadliest terrorist attacks of the decade so far, both attributed to
ISIS. In 2015, both for the abuse-containing tweets and the general
set, the economy is the most prominent topic. National security was
not an important topic in 2015. However, borders and immigration
appears more prominently in the abuse-containing tweets. This may
reflect the fact that EU membership was a key topic in 2015, whereas
by the time of the 2017 election, the 2016 referendum had largely
settled the question.

The absolute numbers above are not large, because abuse-containing
tweets often do not contain an explicit reference to a
topic. Therefore we also analysed the amount of abuse that appeared in
responses to tweets on particular topics. In terms of number of
abusive replies per topic mention, the topic drawing the most abuse in
2017 was national security, at a rate of 0.026 abusive replies per
mention, significantly higher than the mean of 0.014
($p\textless0.0001$, chi-square test with Yates correction). Other
topics particularly drawing abuse, considering only those with at
least 50 abusive replies, are employment (0.019), tax and revenue
(0.019), Scotland (0.019), the UK economy (0.018) and crime and
policing (0.018). Note that the topic replied \textit{to} and the
topic replied \textit{about} may be quite different. For example, a
person might reply to a tweet about the NHS with a contribution on the
subject of immigration.

Finally, we analyzed the topics mentioned across all tweets by
tweeters that sent abuse. We present results for the 2017 time period
only. We collected up to 3,000 posts for each of the 2,506 accounts
that posted at least three abuse-containing messages.  Each of these
accounts was then vectorialized on the basis of topic mentions, one
dimension per topic, and then clustered using k-means. Since the
purpose is exploratory, and the aim is to produce clusters that group
the accounts in a way that facilitates understanding, a k of 8 was
selected, since it makes for a readable result. The following clusters
emerged, described here in terms of their cosine with each topic axis,
giving an indicator of topic satellites that might be used to profile
those that send abuse:

\begin{itemize}
  \item{\textit{The Economy} (409 accounts): Economy
    (0.48), Europe (0.36), public health (0.33), tax (0.24), democracy
    (0.20)}
  \item{\textit{Europe and Trade} (387 accounts): Europe
    (0.85), economy (0.20)}
  \item{\textit{The NHS} (337 accounts): Public health
    (0.63), economy (0.28), Europe (0.25), democracy (0.21), crime
    (0.21)}
  \item{\textit{Borders} (284 accounts): Europe (0.61),
    immigration (0.35), community (0.29), national security (0.22),
    crime (0.22)}
  \item{\textit{Crime} (258 accounts): Crime (0.41),
    national security (0.38), public health (0.26), economy (0.21)}
  \item{\textit{Changing Society} (208 accounts): Community
    (0.63), immigration (0.36), national security (0.31), crime
    (0.26), Europe (0.25)}
  \item{\textit{Scotland} (145 accounts): Scotland (0.71),
    Europe (0.36), economy (0.23)}
  \item{\textit{Schools} (104 accounts): Schools (0.20)}
\end{itemize}

The clusters have intuitively covered a range of viewpoints that might
indicate a strength of opinion. Only cluster 7, ``schools'', is a weak
cluster, and may simply have become a catch-all for accounts with no
strong profile. It shows that despite national security accounting for
the most abuse in 2017, in fact more of those sending abuse are
concerned with issues such as the economy, Europe and the national
health service. National security seems to be the purview of a vocal
minority of those sending abuse.


\section{Discussion and Future Work}

This work provides an empirical contribution to the current debate
around abuse of politicians online, challenging some assumptions
whilst providing a quantified corroboration for others. In particular,
we contribute a detailed investigation into the behaviour and concerns
of those who send abuse to MPs and electoral candidates. Our main
findings are itemized below.

\begin{itemize}

  \item{Abuse received correlates reliably with attention
  received. However, within that, there is a tendency for the more
  prominent politicians to receive proportionally \textit{less} abuse,
  and for those that engage with Twitter to receive more. Male MPs and
  Conservatives were the target of more abuse in the data studied.}

  \item{Abusive behaviour falls into different types. Users who target
  their abuse at a small number of individuals show more evidence of
  using throwaway accounts.}
    
  \item{The topics eliciting abusive tweets differ from those
    discussed by general users. In 2015, borders were a greater topic
    of concern among those sending abuse, whereas in 2017, terrorism
    was.}

  \item{Abuse increased significantly between the 2015 and 2017
  general elections.}

\end{itemize}

Our finding regarding the gender of targeted politicians is in keeping
with the result for the general population reported by Pew Internet
Research.\footnote{http://www.pewinternet.org/2014/10/22/online-harassment/}
They note that whilst men receive more abuse, women are more likely to
be subject to stalking and sexual harrassment and are more likely to
feel upset by it. We did not separate out these types of abuse in the
current study, although it is planned for future work.

Regarding the finding that Conservatives receive more harrassment,
correctly contextualising this might involve contrasting the case
where the Conservatives are not currently in
power. Bartlett\footnote{http://www.telegraph.co.uk/technology/twitter/11400594/Which-party-leader-gets-the-most-abuse-on-Twitter.html}
found the Conservatives were also receiving a more negative response
than Labour Party politicians in February 2015, corroborating our
findings. He further notes, as do Theocharis et
al~\shortcite{theocharis2016bad}, that interactivity, in the sense of
sending fewer broadcast-style tweets and more conversational tweets,
tends to improve the way a politician is perceived. Note that we
didn't differentiate between broadcast and interactive styles of
tweeting in our work. It would be interesting to contextualize our
finding regarding politicians with greater Twitter engagement drawing
more abuse by contrasting the two engagement styles.

The work presented here raises a number of questions about how abuse
is defined and measured, and what it says about the
perpetrator. Wulczyn el al~\shortcite{wulczyn2017ex} note that less than
half of the abuse in Wikipedia talk pages comes from anonymous users,
which is in keeping with our finding that abuse is not something that
necessarily emerges from clearly defined abberant individuals. It also
depends on circumstances, and is part of a broader social
picture. Furthermore, language needs to be contextualised to be
meaningful. A slight on someone's intelligence can be hate speech in
one context and relatively harmless in another. Awan and
Zempi~\shortcite{awan2017will} discuss the particular harm done by hate
speech. Quantitative work agglomerates a number of effects; whilst our
work found no strong evidence of increased abuse towards BAME
politicians, a tendency to show courtesy to minorities generally could
mask a strain of abuse that makes them a particular target.

In future work, we will experiment with more specific classification
of hate speech such as that described by Waseem and
Hovy~\shortcite{waseem2016hateful}. This would allow for further work
focusing specifically on the more severe issue of hate and
intimidation, a dimension vital for understanding the real world
impact of online abuse. It is possible that future work might be able
to quantify the impact that receiving abuse has on a politician;
however, such work would need to be carefully designed, since it would
need to control for the natural rise and fall of a politician's
prominence, and with it the amount of abuse they receive.


\section{Acknowledgments}
\small{
This work was partially supported by the European Union under grant
agreements No. 610829 DecarboNet and 654024 SoBigData, the UK
Engineering and Physical Sciences Research Council (grant
EP/I004327/1), and by the Nesta-funded Political Futures Tracker
project.\footnote{\url{http://www.nesta.org.uk/news/political-futures-tracker}}

\bibliographystyle{aaai}
\bibliography{politicians-twitter-abuse-arxiv}
}
\end{document}